\documentstyle[12pt]{article}
\def \a{\alpha}
\def \b{\beta}
\def \l{\lambda}
\def \d{\delta}
\def \p{\partial}
\def \dg{\dagger}
\def \e{\epsilon}
\def \s{\sigma}
\def \t{\theta}
\def \T{\Theta}

\def \ra{\rightarrow}

\def \am{angular momentum}
\def \fs{fractional spin}
\def \tr{transformation}
\def \g{gauge}
\def \Pm{\pmatrix}

\begin{document}

\centerline{\large{\bf Hamiltonian Analysis of Gauged $CP^1$ Model,}}

\vskip 0.2in

\centerline{\large{\bf the Hopf term, and fractional spin}}

\vskip 0.5in

\centerline{B. Chakraborty\footnote{e-mail: biswajit@boson.bose.res.in}
and A. S. Majumdar\footnote{e-mail: archan@boson.bose.res.in}}

\vskip 0.3in

\centerline{S. N. Bose National Centre for Basic Sciences}

\centerline{Block-JD, Sector-III, Salt Lake, Calcutta-700091, India.}

\vskip 0.8in

 Recently it was shown by Cho and Kimm that the gauged $CP^1$
model, obtained by gauging the	global $SU(2)$ group and adding
a corresponding Chern-Simons term, has got its own soliton. These solitons
are somewhat distinct from those of pure $CP^1$ model as they cannot
always be characterised by $\pi_2(CP^1)=Z$. In this paper, we first
carry out a detailed Hamiltonian analysis of this gauged $CP^1$ model.
This reveals that the model has only $SU(2)$ as the gauge invariance, rather
than $SU(2) \times U(1)$. The $U(1)$ gauge invariance of the original
(ungauged) $CP^1$ model is actually contained in the $SU(2)$ group
itself. Then we couple the Hopf term associated to these solitons and again
carry out its Hamiltonian analysis. The symplectic structures, along
with the structures of the constraints of these two models (with or
without Hopf term) are found to be essentially the same. The model
with a Hopf term is  shown to have fractional spin
which, when computed in the radiation gauge, is found to depend
not only on the soliton number $N$, but also on the nonabelian charge. We
then carry out a reduced (partially) phase space analysis
in a different physical sector of the model
where the degrees of freedom associated with the $CP^1$ fields are
transformed away. The model now reduces to a $U(1)$ gauge theory with
two Chern-Simons gauge fields getting mass-like terms and one remaining
massless. In this case the fractional spin is computed in terms of the
dynamical degrees of freedom and shown to depend purely on the charge of
the surviving abelian symmetry. Although this reduced model is shown to
have its own solitonic configuration, it turns out to be trivial.

\pagebreak

1.{\bf Introduction}

\vskip 0.2in

Recently there has been an upsurge of interest in the study of physics
of $2+1$-dimensional systems. Particularly because of the strange nature
of the Poincare group $ISO(2,1)$ in $2+1$-dimension, in contrast to
$ISO(3,1)$, there arises possibilities of nontrivial configuration space
${\cal Q}$ and the associated fractional spin and statistics. These
possibilities can be realised in practice by adding topological terms like
the Chern-Simons(CS) or Hopf term in the model[1,2]. Fractional spin
and Galilean/Poincare' symmetry in these various
models have been exhibited in detail in the literature, where both path
integral[2,3,4] and the canonical analysis[2,5-9] have been performed.

The CS term (abelian) is a local expression
($\sim \e^{\mu \nu \l}a_{\mu}\p_{\nu}a_{\l}$)
involving a ``photon-less" gauge field $a_{\mu}$[5].
This gauge field is basically introduced to mimic, in the manner of
Aharanov-Bohm, the phase acquired by the system in traversing a
nontrivial loop in the configuration space[2]. On the other hand, the Hopf term
is usually constructed by writing the conserved topological current $j^{\mu}$
$(\p_{\mu}j^{\mu}=0)$ of a model as a curl of a `fictitious' gauge
field $a_{\mu}$:
$$j^{\mu}=\e^{\mu \nu \l}\p_{\nu}a_{\l}\eqno(1.1)$$
and then contracting $j^{\mu}$ with $a_{\mu}$ to get the Hopf term $H$ as,
$$H \sim \int d^3x j^{\mu}a_{\mu}\eqno(1.2)$$
Written entirely in terms of $a_{\mu}$ (using (1.1)), this Hopf term has
also the appearance of CS term. However there is a subtle difference.
In the case of CS term, the gauge field $a_{\mu}$ should be counted as an
independent variable in the configuration space[7-9]. This is despite the fact
that the gauge field is ``photon-less''. On the other hand, the $a_{\mu}$
in (1.1) is really a `fictitious' gauge field (as we have mentioned above)
and is not an independent variable in the configuration space. It
has to be, rather, determined by inverting (1.1), by making use of a suitable
gauge fixing condition. Once done that, the Hopf term (1.2) represents
a non-local current-current interaction. It should however be mentioned
that this distinction, in terminology, has not always been maintained in the
literature (see for example [4]).

It is well known that the Hopf term has geometrical significance in
certain cases.
For example, consider the $O(3)$ non-linear sigma model(NLSM). The model has
solitons[10] characterised by a conserved topological charge $N$. There
exists a topological current $j^{\mu}$ satisfying
$\p_{\mu}j^{\mu}=0$ such that $N = \int d^2x j^0 (x)$.
In this case the Hopf term
provides a representation of the fundamental group of the configuration
space ${\cal Q}(\pi_1({\cal Q}))$. Note that the configuration
space for NLSM is basically given as the space ${\cal Q}=Map (S^2,S^2)$, so
that $\pi_1({\cal Q})=\pi_3(S^2)=Z$. As mentioned earlier, here too
the Hopf term has an inherent
non-locality. Nevertheless, it is possible to write a local version of
the Hopf term in the equivalent
$CP^1$ model[10,11,12] as this is
a $U(1)$ gauge theory having an enlarged phase space.
However it should be mentioned that
this trick of enlarging the phase space and writing a local
expression of the Hopf term may not work all the time, as we shall see later
in this paper.

That the Hopf term can impart fractional spin, was demonstrated initially
by Wilczek and Zee[3], in the context of the NLSM, using path integral
technique. It has been found to depend on the soliton number. This result
was later corroborated by Bowick et al.[13], using canonical quantization.
On the other hand, the fractional spin obtained in the models involving
abelian(nonabelian) CS term, have been found to depend on the total abelian
(nonabelian) charge of the system.

This is an important observation, considering the fact that NLSM has become
almost ubiquitious in physics, appearing in various circumstances where the
original $O(3)$ symmetry is broken spontaneously. For example, in particle
physics, the model is considered a prototype of QCD, as the model is
asymptotically free in $(1+1)$ dimension. On the other hand, in condensed
matter physics, this model can describe antiferromagnetic spin chain in its
relativistic version[14]. And in its nonrelativistic version, it can describe a
Heisenberg ferromagnetic system in the long wavelength limit, i.e. the
Landau-Lifshitz(LL) model[15,16]. Besides, the Hopf term can arise naturally in this
NLSM, when one quantises a $U(1)$ degree of freedom hidden in the configuration
space ${\cal Q}$, as has been shown recently by Kobayashi et. al.[17]. Further,
it has been shown recently in [12], that the Hopf term can alter the spin
algebra of the LL model drastically.

This NLSM has global $O(3)$ symmetry. Recently Nardelli[18] has shown that
if this $O(3)$ group is gauged by adding an $SO(3)$ CS term, then the resulting
model also has got its own soliton. This work was later extended by Cho and
Kimm[19] for the general $CP^N$ model, where one has to gauge the global
$SU(N+1)$ group and add a corresponding CS term.
These solitons are somewhat distinct from those of pure
$CP^N$ model, in the sense that these are not always characterised by the second
homotopy group ($\pi_2(CP^N)=Z$) of the manifold, unlike the pure $CP^1$
model[10].

The purpose of the paper is to investigate ($N=1$ case), whether a Hopf term
associated with this new soliton number can be added to the model to
obtain fractional spin. The question is all the more important, as the model
has already got an nonabelian $SU(2)$ CS term, needed for the {\it very existence}
of these new type of solitons. And, as we have mentioned earlier, the CS
term, is likely to play its own role in imparting fractional spin to the
model. We find that the fractional angular momentum is given in terms of
both the soliton number and the nonabelian charge, where the radiation
gauge condition is used. This corresponds to one particular sector of
the  theory. Apart from this sector, one can also
consider a different physical sector in which all the degrees of freedom
associated with the $CP^1$ variables are transformed away using the
local $SU(2)$ gauge transformation. The physical differences between these two
sectors arise from the fact that the model is invariant under only those
gauge transformations which tend to a constant at infinity (the
spacetime infinity is mapped into one point on the group manifold).
For example, because of the constraint on the
magnitude of the $CP^1$ variables, one can expect the CS gauge fields to
acquire mass-like terms {\it a la} the Higgs mechanism in the standard
model. This, on turn, is expected to affect the asymptotic proprties of
the fields, with non-trivial effects on the physical observables,
particularly, the \am. We find that one $U(1)$ gauge symmetry survives
in the Lagrangian of this partially reduced configuration space, and the
{\am} is now given in terms of the abelian charge. This shows that the
computation of fractional spin yields different results in different
physical sectors associated with different asymptotic behaviour of the gauge
variant fields. Lastly, we shall investigate the role of the Hopf term
in this latter \g. In doing so, we find that although this model admits
a static minimum energy configuration, the solitonic charge in this case
is given by the Noether charge, with vanishing Hopf term.

To that end, we organise the paper as follows. In section 2, we carry out
the Hamiltonian analysis of the gauged $CP^1$ model. The Hopf term is introduced
in section 3 and again the Hamiltonian analysis of the resulting model is
performed. In section 4, we compute the fractional spin of the model.
We carry out a reduced phase space analysis corresponding to a different
physical sector of the theory in section 5 and calculate
the {\fs} in terms of the surviving dynamical degrees of freedom.
Finally we conclude in section 6.

\vskip 0.2in

2.{\bf Hamiltonian Analysis of Gauged $CP^1$ model}

\vskip 0.1in

We are going to carry out the Hamiltonian analysis of the gauged
$CP^1$ model, as introduced by Cho and Kimm[19]. The model is
given by,
$${\cal L}=(D_\mu Z)^{\dg}(D^{\mu}Z)+\t \e^{\mu \nu \l}
[{A^a}_{\mu}\p_{\nu}{A^a}_{\l}+{g\over 3}\e^{abc}A^a_{\mu}A^b_{\nu}A^c_{\l}]
-{\l}(Z^{\dg}Z-1)\eqno(2.1)$$
where $Z=\pmatrix{z_1 \cr z_2}$ is an $SU(2)$ doublet satisfying
$$Z^{\dg}Z=1\eqno(2.2)$$
and enforced by the Lagrange multiplier $\l$ in (2.1). The
covariant derivative operator $D_{\mu}$ is given as
$$D_{\mu}=\p_{\mu}-ia_{\mu}-igA^a_{\mu}T^a \eqno(2.3)$$
with $T^a={1\over 2}{\s}^a$ (${\s}^a$, the Pauli matrices), representing
the $SU(2)$ generators and $g$ a coupling constant.$\t$ represents the
CS parameter.And finally $a_{\mu}$ and $A_{\mu}^a$ represent the
$U(1)$ and $SU(2)$ gauge fields respectively. Note that there is no
dynamical CS term for the the $a_{\mu}$ field.

The canonically conjugate momenta variables corresponding to the
configuration space variables ($a_{\mu}$,$A_{\mu}^a$,$z_{\a}$,$z_{\a}^*$)
are given as,
$$\pi^{\mu}={\d L\over {\d \dot {a_{\mu}}}}=0$$
$$\Pi^{ia}={\d L\over {\d \dot {A^a_i}}}=\t \e^{ij}A^a_j;
\Pi^{0a}={\d L\over {\d \dot {A^a_0}}}=0\eqno(2.4)$$
$$\pi_{\a}={\d L\over {\d \dot z_{\a}}}=(D_0Z)^*_{\a};
\pi^*_{\a}={\d L\over {\d \dot z_{\a}^*}}=(D_0Z)_{\a}$$
where $L=\int d^2x {\cal L}$ is the Lagrangian.

The Legendre transformed Hamiltonian
$${\cal H}=\pi^{\mu}{\dot a_{\mu}}+\Pi^{\mu a}{\dot A^a_{\mu}}+\pi_{\a}
{\dot z}_{\a}+\pi_{\a}^*{\dot z}^*_{\a}-\cal L \eqno(2.5a)$$
when expreseed in terms of the phase space variables (2.4), gives
$${\cal H}=\pi_{\a}^*\pi_{\a} +ia_0(\pi_{\a}z_{\a}-\pi_{\a}^*z_{\a}^*)+
{1\over 2}igA^a_0[\pi_{\a}(\s^aZ)_{\a}-\pi^*_{\a}(Z^{\dg}\s^a)_{\a}]$$
$$-2 \t A^a_0B^a+(D_iZ)^{\dg}(D_iZ)+\l (Z^{\dg}Z-1)\eqno(2.5b)$$
where
$$B^a\equiv F^a_{12}=(\p_1A_2^a-\p_2A^a_1+g\e^{abc}A_1^bA_2^c)\eqno(2.5c)$$
is the non-abelian $SU(2)$ magnetic field.

Clearly the fields $a_0$,$A^a_0$ and $\l$ play the role of Lagrange
multipliers, which enforce the following constraints,
$$G_1(x) \equiv i(\pi_{\a}(x)z_{\a}(x)-\pi_{\a}^*(x)z_{\a}^*(x))\approx 0\eqno(2.6)$$
$$G_2^a(x) \equiv {ig \over 2}[\pi_{\a}(x)(\s^aZ(x))_{\a}-\pi^*_{\a}(x)(Z^{\dg}(x)\s^a)_{\a}]-2 \t B^a(x) \approx 0\eqno(2.7)$$
$$\chi_1(x) \equiv Z^{\dg}(x)Z(x)-1 \approx 0.\eqno(2.8)$$
Apart from all these constraints, we have yet another primary constraint,
$$\chi_2(x) \equiv (\pi_{\a}(x)z_{\a}(x)+\pi_{\a}^*(x)z_{\a}^*(x))\approx 0\eqno(2.9)$$
Also the preservation of the primary constraint $\pi^i(x) \equiv 0$ (2.4)
yield the following secondary constraint,
$$2iZ^{\dg}\p_jZ + 2a_j + gA_j^aM^a \approx 0 \eqno(2.10)$$
Here
$$M^a=Z^{\dg}\s^aZ\eqno(2.11)$$
is a unit 3-vector, obtained from the $CP^1$ variables using the Hopf map.
We are left with a pair of primary constraints from the CS gauge field
sector in (2.4),
$$\xi^{ia}\equiv \Pi^{ia}-\t \e^{ij}A^a_j\approx 0\eqno(2.12)$$
This pair of constraints can be implemented strongly by the bracket,
$$\{A^a_i(x),A^b_j(y)\}={1\over 2 \t}\e_{ij}\d^{ab}\d (x-y)\eqno(2.13)$$
obtained either by using
Dirac method[20] or by the symplectic technique of Faddeev-Jackiw[21].

Also note that the constraint $\pi^i\approx 0$ (2.4) is conjugate to the
constraint (2.10) and can again be strongly implemented by the Dirac bracket
(DB),
$$\{\pi^i(x),a_j(y)\}=0\eqno(2.14)$$
With this the `weak' equality in (2.10) is actually rendered into a strong
equality and the field $a_i$ ceases to be an independent degree of freedom.

Finally the constraints $\chi_1$ (2.8) and $\chi_2$ (2.9) are conjugate
to each other and are implemented strongly by the following DBs,
$$\{z_{\a}(x),z_{\b}(y)\}= \{z_{\a}(x),z_{\b}^*(y)\}=0$$
$$\{z_{\a}(x),\pi_{\b}(y)\}=(\d_{\a \b}-{1\over 2}z_{\a}z_{\b}^*)\d (x-y)$$
$$\{z_{\a}(x),\pi_{\b}^*(y)\}=-{1\over 2}z_{\a}z_{\b}\d (x-y)\eqno(2.15)$$
$$\{\pi_{\a}(x),\pi_{\b}(y)\}=-{1\over 2}(z_{\a}^*\pi_{\b}-z_{\b}^*\pi_{\a})\d (x-y)$$
$$\{\pi_{\a}(x),\pi_{\b}^*(y)\}=-{1\over 2}(z_{\a}^*\pi_{\b}^*-z_{\b}\pi_{\a})\d (x-y)$$
Precisely the same set of brackets (2.15) are obtained in the case of $CP^1$
model also[22]. We are thus left with the constraints (2.6) and (2.7) and are
expected to be the Gauss constraints generating $U(1)$ and $SU(2)$ gauge
transformations respectively. The fact that this is indeed true will be
exhibited by explicit computations. But before we proceed further, let
us note that the constraints (2.8),(2.9) and (2.10) hold strongly now. In
view of this, the constraint $G_1$ (2.6) can be simplified as,
$$G_1(x)=2i\pi_{\a}(x)z_{\a}(x) \approx 0 \eqno(2.16)$$
At this stage, one can substitute $\pi_{\a}=(D_0Z)^{\dg}_{\a}$ from (2.4)
and solve for $a_0$ to get,
$$a_0=-iZ^{\dg}\p_0Z-{1\over 2}gA^a_0M^a \eqno(2.17)$$
Clearly this is not a constraint equation, as it involves time derivative.
It is nevertheless convenient to club it with the expression of $a_i$,
obtained from (2.10) and write covariantly as,
$$a_{\mu}=-iZ^{\dg}\p_{\mu}Z-{1\over 2}gA^a_{\mu}M^a \eqno(2.18)$$
Here the first term $(-iZ^{\dg}\p_{\mu}Z)$ is the pullback, onto the
spacetime, of the $U(1)$ connection on the $CP^1$ manifold[16]. The second
term on the other hand has nothing to do with $CP^1$ connection and arises
from the presence of the CS gauge field $A^a_{\mu}$.

It is now quite trivial to show that $G_1(x)$ (2.16) generates $U(1)$ gauge
transformation on the $Z$ fields
$$\d Z(x)=\int d^2y f(y)\{Z(x),G_1(y)\}=if(x)Z(x)\eqno(2.19)$$
but leaves the CS gauge field $A^a_{\mu}$ unaffected
$$\d A^a_{\mu}(x)=\int d^2y f(y)\{A^a_{\mu}(x),G_1(y)\}=0\eqno(2.20)$$
Consequently $M^a=Z^{\dg}\s^aZ$ (2.11) remains invariant under this
transformation and hence $a_{\mu}$ (2.18) undergoes the usual gauge
transformation
$$\d a_{\mu}(x)=\int d^2y f(y)\{a_{\mu}(x),G_1(y)\}=\p_{\mu}f(x) \eqno(2.21)$$
Here in the equations (2.19-2.21) we have taken $f(x)$ to be an arbitrary
differentiable functions with compact support.

Proceeding similarly, one can show that the constraints $G_2^a(x)$ (2.7)
generates $SU(2)$ gauge transformation,
$$\d Z(x)\equiv \int d^2y f^a(y)\{Z(x),G^a_2(y)\}=ig f^a(x)(T^aZ(x))\eqno(2.22a)$$
$$\d A^a_i(x)\equiv \int d^2y f^b(y)\{A^a_i(x),G^b_2(y)\}=
\p_if^a(x)-g\e^{abc}f^b(x)A^c_i(x)\eqno(2.22b)$$
Using these one can also show that,
$$\d B^a=-g\e^{abc}f^bB^c$$
$$\d M^a=-g\e^{abc}f^bM^c\eqno(2.23a)$$
but $(M^aB^a)$ is an $SU(2)$ scalar as
$$\d (M^aB^a)=0\eqno(2.23b)$$
It also follows from (2.23a) and (2.18) that $a_{\mu}$ remains unaffected
by this $G_2^a$,
$$\d a_{\mu}(x)=\int d^2y f^a(y)\{a_{\mu}(x),G^a_2(y)\}=0\eqno(2.24)$$
just as $A_{\mu}^a(x)$ remains unaffected by $G_1$ (2.20).

The fact that $G_1(x)$ and $G_2^a(x)$ are indeed  the first class constraints
of the model can be easily seen. Firstly one has to just rewrite $G_1$(2.16)
using (2.4) as
$$G_1(x)=2i(D_0Z)^{\dg}Z \approx 0\eqno(2.25)$$
to see that this is manifestly invariant under the $SU(2)$ gauge transformation
generated by $G_2^a(x)$ (2.22). We thus have,
$$\{G_1(x),G^a_2(y)\}=0\eqno(2.26)$$
It also follows after a straightforward algebra that $G^a_2$'s satisfy an
algebra isomorphic to $SU(2)$ Lie algebra and thus vanishes on the constraint
surface,
$$\{G^a_2(x),G^b_2(y)\}=2\e^{abc}G_2^c(x)\d (x-y)\approx 0\eqno(2.27)$$

At this stage one can observe that the infinitesimal {\g} {\tr} generated by
$G_1$ (2.19) can be integrated to get the following \tr
$$Z(x) \ra Z'(x) = e^{if(x)} Z(x)\eqno(2.28)$$
Here $f$ is taken to be a finite quantity. Although such a {\tr} matrix
$\pmatrix{e^{if} & 0 \cr 0 & e^{if}} \in U(2)$ is not an element of
$SU(2)$, it nevertheless generates an orbit in $S^3$, taking a point
$Z= \pmatrix{z_1 \cr z_2}$ on $S^3$ to another point $e^{if}Z$ on the
same manifold. On the other hand, we also know that the group $SU(2)$
acts transitively on $S^3$. Indeed, one can check that the following
$SU(2)$ action on $Z$ is identical to the $U(1)$ \tr (2.28):

$$\Pm{z_1 \cr z_2} \ra \Pm{a & b \cr -b^* & a^*} \Pm{z_1 \cr z_2} =
e^{if} \Pm{z_1 \cr z_2}\eqno(2.29)$$
with
$$\Pm{a \cr b} = \Pm{\vert z_1\vert^2 e^{if} + \vert z_2\vert^2 e^{-if}
\cr z_1z^*_2 e^{if} - z_1z^*_2 e^{-if}}\eqno(2.30)$$
so that $\Pm{a & b \cr -b^* & a^*} \in SU(2)$. Considering again the
case where $|f(x)| \ll 1$, one can get an element
$$\Pm{(1+ifM_3) & if(M_1-iM_2) \cr if(M_1+iM_2) & (1-ifM_3)} =
\Biggl[\Pm{1 & 0 \cr 0 & 1} + ifM^a\sigma^a\Biggr] \in
SU(2)\eqno(2.31)$$
close to the identity of $SU(2)$. The associated $SU(2)$ Lie algebra
element is therefore $fM^a\sigma^a$. On the other hand, the
corresponding $U(1)$ Lie algebra element, from (2.28), is simply $f{\cal
I}$, where ${\cal I}$ is the $2\times 2$ unit matrix.
Since the Gauss constraints $G^a_2$ and $G_1$ satisfy algebra isomorphic
to the $SU(2)$ and $U(1)$ Lie algebra (2.26, 2.27) respectively, one can
expect the following relation
$$G_1(x) = M^a(x) G^a_2(x)\eqno(2.32)$$
to hold. To prove that this is indeed the case, first consider the
quantity $M^aB^a$ which we have shown to be an $SU(2)$ scalar (2.23b).
Therefore, it can be evaluated in any gauge of our choice. Choosing $Z = \Pm{0
\cr 1}$, we get $M^a = -\delta^{a3}$. Thus, the constraint $G^3_2
\approx 0$ (2.7) reduces to $B^3 \approx 0$, and it follows that
$$M^aB^a = 0\eqno(2.33)$$
Now using the definitions of $\pi_{\a}$ and
$\pi^*_{\a}$ (2.4) in (2.7), it is easy to show that (2.32) is
satisfied. This shows that $G_1$ (2.6) is not an independent constraint.
In other words, the $U(1)$ symmetry {\tr} generated by (2.6) is not a
different kind of {\tr}, but can be generated by a suitable combination
(2.32) of
$G^a_2$ (2.7) itself. Thus, strictly speaking the model has only an
$SU(2)$ {\g} invariance. In this context, it is useful to distinguish the
{\tr} generated by this $U(1)$ from the {\tr} generated by the $U(1)$
subgroup of $SU(2)$. The latter acts as
$$Z \ra Z' = \Pm{e^{if} & 0 \cr 0 & e^{-if}}Z\eqno(2.34)$$
and the corresponding generator is $G^3_2$.

Finally note that (2.18) really corresponds to the Euler-Lagrange's equation
for the $a_{\mu}$ field. The corresponding equations for $Z$ and $A^a_{\l}$
are given by,
$$D_{\mu}D^{\mu}Z + \l Z =0 \eqno(2.35)$$
$$\t \e^{\mu \nu \l}F_{\nu \l}^a =
ig[(D^{\mu}Z)^{\dg}T^aZ- Z^{\dg}T^a(D^{\mu}Z)]\eqno(2.36)$$
respectively.

\vskip 0.2in

3.{\bf Introducing the Hopf term}

\vskip 0.1in

In order to introduce the Hopf term, it will be convenient to provide
a very brief review of some of the essential features of these new solitons.
For this we essentially follow [19].
The symmetric expression for the energy-momentum(EM) tensor, as obtained by
functionally differentiating the action $S(=\int d^3x {\cal L})$ with respect
to the metric, is given by
$$T_{\mu \nu}=(D_{\mu}Z)^{\dg}(D_{\nu}Z)+(D_{\nu}Z)^{\dg}(D_{\mu}Z)
-g_{\mu \nu}(D_{\rho}Z)^{\dg}(D^{\rho}Z)\eqno(3.1)$$
The energy functional
$$E=\int d^2x T_{00}=\int d^2x [2(D_0Z)^{\dg}(D_0Z)-(D_{\mu}Z)^{\dg}(D^{\mu}Z)]\eqno(3.2)$$
can be expressed alternatively as,
$$E=\int d^2x ( |D_0Z|^2+ |(D_1\pm iD_2)Z|^2) \pm 2\pi N \eqno(3.3a)$$
where
$$N={1\over 2\pi i}\int d^2x \e^{ij}(D_iZ)^{\dg}(D_jZ)\eqno(3.3b)$$
is the soliton charge.

It immediately follows that the energy functional satisfy the following
inequality,
$$E \geq 2\pi |N| \eqno(3.4)$$
The corresponding saturation conditions are,
$$|D_0Z|^2=|(D_1\pm iD_2)Z|^2=0 \eqno(3.5)$$
For static configuration ($\dot Z=0$), this yields
$$A^a_0=k M^a\eqno(3.6)$$
where $k$ is an arbitrary constant.

Again assuming the static case, one can easily show that $\mu =0$ component
of the Euler-Lagrange equation (2.29) implies that the $SU(2)$ magnetic
field $B^a$ vanishes,
$$B^a=0 \eqno(3.7)$$
where use of (3.6) has been made. This in turn implies that $A^a_i$ is a
pure gauge, so that one can write without loss of generality
$$A^a_i=0 \eqno(3.8)$$
In this gauge, the soliton charge $N$ (3.3b) reduces to the standard $CP^1$
soliton charge,
$$N={1\over 2\pi i}\int d^2x \e^{ij}({\cal D}_iZ)^{\dg}({\cal D}_jZ)\eqno(3.9a)$$
where
$${\cal D}_i=D_i|_{A_i^a=0}=\p_i-(Z^{\dg}\p_iZ)\eqno(3.9b)$$
is the covariant derivative operator for the standard $CP^1$ model. Thus
in this gauge (3.8), the ``soliton charge" is essentially characterised by
$\pi_2(CP^1)=Z$. Nonetheless, it is possible to make ``large'' topology
changing gauge transformation, where $A^a_i$ is no longer zero and one
{\it has} to make use of (3.3b), rather than (3.9a), to compute the
solitonic charge. Of course this will yield the same value for the charge,
but the various solitonic sectors will not be characterised by $\pi_2(CP^1)$
anymore.

To make things explicit, consider a typical solitonic configuration:
$$Z={1\over {\sqrt {r^2+\l^2}}}\Pm{re^{-i\Phi} \cr \l}\eqno(3.10a)$$
$$A_i^a=0\eqno(3.10b)$$
where ($r$,$\Phi$) represents the polar coordinates in the two-dimensional
plane and $\l$ is the size of the soliton. The corresponding unit vector
$M^a$(2.11) takes the form,
$$M^1=sin \T cos \Phi ={2r\l \over {r^2+ \l^2}}cos \Phi$$
$$M^2=sin \T sin \Phi ={2r\l \over {r^2+ \l^2}}sin \Phi \eqno(3.11)$$
$$M^3=cos \T ={r^2-\l^2 \over {r^2+\l^2}}$$
We therefore have for the time component of the gauge field $A_0^a=kM^a$(3.6).

At this stage, one can make a topology changing	 transformation,
$$Z \rightarrow Z'=UZ=\Pm{0 \cr 1}\eqno(3.12a)$$
where,
$$U={1\over {\sqrt {r^2+\l^2}}}\pmatrix{\l & -re^{-i\Phi} \cr re^{i\Phi} & \l}\in SU(2)\eqno(3.12b)$$
so that $A^a_0$ undergoes the transformation,
$$A^a_0\rightarrow {A'}^a_0=-k\pmatrix{0 \cr 0 \cr 1}\eqno(3.12c)$$
The spatial components on the other hand, undergoes the transformation,
$$A_i\rightarrow A'_i=UA_iU^{-1}+{i\over g}U\p_iU^{-1}=-{i\over g}(\p_iU)U^{-1}$$
which on further simplification yields the following form for the connection
one-form in the cartesian coordinate system,
$${A'}^1={2\l \over g(r^2+\l^2)}dy$$
$${A'}^2=-{2\l \over g(r^2+\l^2)}dx\eqno(3.12d)$$
$${A'}^3=-{2 \over g(r^2+\l^2)}(xdy-ydx)$$
Now it is a matter of straightforward exercise to calculate the soliton charge
`$N$' in either of these gauges (3.10) and (3.12). For example, in the gauge
(3.10), this can be computed by using (3.9a) to get,
$$N={1\over {2\pi}}\int d(-iZ^{\dg}dZ)=-1 \eqno(3.13)$$
On the other hand, the same soliton charge can also be computed in the gauge
(3.12), but where the use of (3.3b), rather than (3.9a), has to be made. Note
that the topological density $j^0$ ($N\equiv \int d^2x j^0$) can be written as,
$$j^0={1\over 2\pi i}\e^{ij}(D_iZ)^{\dg}(D_jZ)={\tilde j}^0+{g\over 4\pi}
\e^{ij}A^a_i(\p_jM^a+{g\over 2}\e^{abc}A^b_jM^c)\eqno(3.14a)$$
where,
$${\tilde {j}}^0={\e^{ij}\over 2\pi i}({\cal D}_iZ)^{\dg}({\cal D}_jZ)\eqno(3.14b)$$
is the expression of topological density in the gauge (3.10). But in the gauge (3.12),
this ${\tilde j}^0$ vanishes, and one can rewrite $N$ completely in terms of the
CS gauge field as,
$$N={g^2\over {8\pi k}}\int d^2x \e^{ij} \e^{abc}A^a_0A^b_iA^c_j \eqno(3.15)$$
The corresponding $Z$ field configuration being trivial ($Z=\pmatrix{0 \cr 1}$),
the soliton number $N$ cannot be captured by $\pi_2(CP^1)$.
Incidentally, we shall see in section $5$ that it is possible to write a
reduced form of the model (2.1) by going to the gauge $Z = \Pm{0 \cr
1}$. However, the corresponding solitons turn out to be trivial with
zero solitonic charge.

At this point we wish to make certain clarifications. The $SU(2)$
transformation (3.12b) which takes the $Z$-field configuration (3.10) to
to $Z=\pmatrix{0 \cr 1}$  (3.12a) does not tend to a constant
asymptotically and hence does not belong to the gauge
group  of the model
(2.1). As mentioned earlier, because of the presence of the $SU(2)$ CS
term in (2.1), the gauge group of this model consists of only those
elements of $SU(2)$ which become constant asymptotically. The
configurations (3.10) and (3.12) are therefore not physically equivalent
and represent different physical states, even though both of them are
associated with the same solitonic charge. In other words, the
transformation (3.12) is, strictly speaking, not a gauge transformation,
but rather a transformation that connects two different physical sectors
of the theory.

Since `$N$' is a conserved soliton charge, with an associated topological
density $j^0$(3.14a), one can regard $j^0$ to be the time-component of a
conserved topological $3(=2+1)$-current
$$j^{\mu}={1\over {2\pi i}}\e^{\mu \nu \l}(D_{\nu}Z)^{\dg}(D_{\l}Z)\eqno(3.16)$$
This can therefore be expressed as the curl of a `fictitious' $U(1)$
gauge field ${\cal A}_{\l}$:
$$j^{\mu}={1\over {2\pi}}\e^{\mu \nu \l}\p_{\nu}{\cal A}_{\l}\eqno(3.17)$$
Unlike the case of pure $CP^1$ model[11,12], this equation cannot be solved
trivially for ${\cal A}_{\l}$ in a gauge independent manner[2,16]. We therefore
find it convenient to follow Bowick et.al.[13], to solve (3.17) for ${\cal A}_{\l}$
in the radiation gauge ($\p_i{\cal A}_i=0$), where one can prove the
following identity,
$$\int d^3x j_0{\cal A}_0=- \int d^3x j_i{\cal A}_i\eqno(3.18)$$
so that the Hopf action
$$S_{Hopf}=\T \int d^3x j^{\mu}{\cal A}_{\mu}, \eqno(3.19)$$
($\T $ being the Hopf parameter and should not be confused with the spherical
angles introduced in (3.11)) simplifies to the following non-local term
$$S_{Hopf}=-2 \T \int d^3x j_i{\cal A}_i. \eqno(3.20)$$
Adding this term to the original model (2.1), we get the following model,
$${\cal L}=(D_\mu Z)^{\dg}(D^{\mu}Z)+\t \e^{\mu \nu \l}
[{A^a}_{\mu}\p_{\nu}{A^a}_{\l}+{g\over 3}\e^{abc}A^a_{\mu}A^b_{\nu}A^c_{\l}]$$
$$+{\T \over \pi i}\e^{ij}{\cal A}_i[(D_jZ)^{\dg}(D_0Z)-(D_0Z)^{\dg}(D_jZ)]
 -{\l}(Z^{\dg}Z-1)\eqno(3.21)$$
In the rest of this section, we shall be primarily concerned with
the Hamiltonian analysis of this model. As the Hopf term is linear in time
derivative of the $Z$ variable, the analysis is expected to undergo only minor
modification. Indeed we shall verify this by explicit computations.

To begin with, note that the only change in the form of canonically conjugate
momenta variables takes place in the variables $\tilde {\pi}_{\a}$ and its
complex conjugates, counterpart of  $\pi_{\a}$ and $\pi_{\a}^*$ (2.4)-the
momenta variables for the model (2.1). They are now given as,
$$\tilde {\pi}_{\a}=(D_0Z)^*_{\a}+{\T \over {\pi i}}\e^{ij}{\cal A}_i(D_jZ)^*_{\a}
=\pi_{\a}+{\T \over {\pi i}}\e^{ij}{\cal A}_i(D_jZ)^*_{\a}$$
$$\tilde {\pi}^*_{\a}=(D_0Z)_{\a}-{\T \over {\pi i}}\e^{ij}{\cal A}_i(D_jZ)_{\a}
=\pi^*_{\a}-{\T \over {\pi i}}\e^{ij}{\cal A}_i(D_jZ)_{\a}\eqno(3.22)$$
 Rest of the momenta variables undergo no change from that
of (2.4).

The Legendre transformed Hamiltonian $\tilde {\cal H}$ can be calculated in
a straightforward manner to get,
$$\tilde {\cal H}={\cal H}+{g\T \over 2\pi}A^a_0\e^{ij}{\cal A}_i(D_jM)^a\eqno(3.23)$$
where $\cal H$ is just the expression of the Legendre transformed Hamiltonian
density (2.5) corresponding to the model (2.1) and $(D_jM)^a$ is given by,
$$D_jM^a=\p_jM^a+g\e^{abc}A^b_jM^c\eqno(3.24)$$
as can be easily obtained by using the Hopf map (2.11) and the fact that the
covariant derivative operator $D_{\mu}$ boils down, using (2.3) and (2.18) to,
$$D_{\mu}Z=\p_{\mu}Z-(Z^{\dg}\p_{\mu}Z)Z+{ig \over 2}A^a_{\mu}(M^a-\s^a)Z\eqno(3.25)$$
Clearly the structure of all the constraints remain the same, except the $SU(2)$
Gauss constraint. This is clearly given as,
$${\tilde G}^a_2=G^a_2+{g\T \over 2\pi}\e^{ij}{\cal A}_i(D_jM)^a \eqno(3.26)$$
where $G^a_2$ is given in (2.7). But we have to rewrite this in terms of
${\tilde \pi}_{\a}$ and ${\tilde \pi}^*_{\a}$. Once we do this, we find that
that the $SU(2)$ Gauss constraint $G^a_2$ (2.7) for the model (2.1) is now
given by,
$$G^a_2=ig\Bigl([{\tilde {\pi}}_{\a}(T^aZ)_{\a}-{\tilde {\pi}}^*_{\a}(Z^{\dg}T^a)_{\a}]
+{i\T \over {2\pi}}\e^{ij}{\cal A}_i(D_jM)^a\Bigr)-2\t B^a\approx 0$$
Substituting this in (3.26), ${\tilde G}^a_2$ is found to have the same form
as that of $G^a_2$ (2.7) with the replacement $\pi_{\a}\rightarrow {\tilde \pi}_{\a}$
and $\pi^*_{\a}\rightarrow {\tilde \pi}^*_{\a}$,
$${\tilde G}^a_2=ig({\tilde {\pi}}_{\a}(T^aZ)_{\a}
-{\tilde {\pi}}^*_{\a}(Z^{\dg}T^a)_{\a})-2\t B^a\approx 0\eqno(3.27)$$
The other $U(1)$ Gauss constraint ${\tilde G}_1$ (2.6) can also be seen to take
the same form, with identical replacement,
$${\tilde G}_1(x)=i({\tilde \pi}_{\a}(x)z_{\a}(x)
-{\tilde \pi}^*_{\a}{z}^*_{\a}(x))\approx 0\eqno(3.28)$$
where use of the identity $Z^{\dg}D_{\mu}Z=0$ has been made. Here again,
one can show that the relation ${\tilde G}_1 = M^a {\tilde G}^a_2$ (the
counterpart of (2.33) holds. This shows that the only basic first class
constraints are given by ${\tilde G}^a_2$ (3.27).

Finally note that the constraint (2.9) also preserves its form,i.e.
$$\chi_2(x)={\tilde \pi}_{\a}(x)z_{\a}(x)+ c.c $$
and is again conjugate to $\chi_1$ (2.8). Thus these pair of constraints
can be implemented strongly by using the DB (2.15), again taken with the
replacement $\pi_{\a}\rightarrow {\tilde \pi}_{\a}$ and
$\pi^*_{\a}\rightarrow {\tilde \pi}^*_{\a}$. On the other hand, the pair
of second class constraints (2.12) are implemented strongly by the brackets
(2.13) in this case also. These set of DB furnishes us with the symplectic
structure of the model (3.21).

\vskip 0.2in

4. {\bf Angular momentum}

\vskip 0.1in

In this section, we are going to find the fractional spin imparted by
the Hopf term. As was done for the models involving the CS[7-9] and Hopf[13]
term, the fractional spin was essentially revealed by computing the
difference $(J^s-J^N)$ between the expression of angular momentum $J^s$,
obtained from the symmetric expression of the EM tensor $T_{\mu \nu}$
($\sim {\d S\over \d g^{\mu \nu}}$) and the one $J^N$, obtained by using
Noether's prescription. It is $J^s$, which is taken to be the physical
angular momentum. This is because it is gauge invariant by construction,
in contrast to $J^N$, which turn out to be gauge invariant only on the
Gauss constraint surface and that too usually under those gauge transformations,
which tend to identity asymptotically[8,9].

To that end, let us consider the generator of linear momentum. This is
obtained by integrating the (0i) component of the EM tensor (3.1), which
undergoes no modification as the metric independent topological (Hopf)
term (3.20) is added to the original Lagrangian (2.1) to get the model (3.21).
$$P_i^s=\int d^2x T^s_{0i}=\int d^2x[(D_0Z)^{\dg}(D_iZ)+(D_iZ)^{\dg}(D_0Z)]\eqno(4.1)$$
Expressing this in terms of phase-space variables (3.22), one gets
$$P_i^s=\int d^2x[{\tilde \pi}_{\a}(D_iZ)_{\a}+{\tilde \pi}^*_{\a}(D_iZ)^*_{\a}
+2 \T{\cal A}_i(x)j^0(x)]\eqno(4.2)$$
This can now be re-expressed as,
$$P_i^s=\int d^2x [{\tilde \pi}_{\a}\p_iz_{\a}+{\tilde \pi}^*_{\a}\p_iz^*_{\a}
-2\t A^a_iB^a+2\T{\cal A}_ij^0-a_i{\tilde G}_1-A^a_i{\tilde G}^a_2]\eqno(4.3)$$
However this cannot be identified as an expression of linear momentum, as this
fails to generate appropriate translation,
$$\{Z(x),P^s_i\}\approx D_iZ \eqno(4.4)$$
in contrast to the corresponding expression of linear momentum
$$P_k^N=\int d^2x T^N_{0k}=\int d^2x [{\tilde \pi}_{\a}\p_kz_{\a}
+{\tilde \pi}^*_{\a}\p_kz^*_{\a}-\t \e^{ij}A^a_i\p_kA^a_j]\eqno(4.5)$$
obtained through Noether's prescription, as this generates appropriate
translation by construction,
$$\{Z(x),P^N_k\}=\p_kZ(x)$$
$$\{A^a_i(x),P^N_k\}=\p_kA^a_i\eqno(4.6)$$
The adjective ``appropriate'' in this context means that the bracket
$\{\Phi (x),{\cal G}\}$ is just equal to the Lie derivative
(${\cal L}_{V_{\cal G}}({\Phi (x)})$) of a generic field $\Phi (x)$
with respect to the vector field $V_{\cal G}$, associated to the symmetry
generator ${\cal G}$. We have not,of course, displayed any indices here.
The field $\Phi$ may be a scalar, spinor, vector or tensor field in
general. In this case, it
can correspond either to the scalar field $Z(x)$ or the vector field
$A^a_i(x)$.
And ${\cal G}$ can be, for example, the momentum($P_i$) or angular momentum
($J$)operator generating translation and spatial rotation respectively. The
associated vector fields $V_{\cal G}$ are thus	given as $\p_i$ and $\p_{\phi}$
respectively ($\phi$ being the angle variable in the polar coordinate system
in 2-dimensional plane).

Coming back to the translational generator $P_i^s$ (4.3), we observe that
the EM tensor (3.1) is not unique by itself. One has the freedom to modify
it to ${\tilde T}_{\mu \nu}$ by a linear combination of first class constraint(s), here
${\tilde G}_1$ (3.28) and ${\tilde G}^a_2$ (3.27) with arbitrary tensor valued
coefficients $u_{\mu \nu}$ and $v^a_{\mu \nu}$:
$${\tilde T}_{\mu \nu}=T_{\mu \nu}+u_{\mu \nu}{\tilde G}_1
+v^a_{\mu \nu}{\tilde G}^a_2\eqno(4.7)$$
Choosing,
$$u_{0i}=a_i$$
$$v^a_{0i}=A^a_i \eqno(4.8)$$
one can easily see that the corresponding modified expression of momentum
$${\tilde P}_i=\int d^2x {\tilde T}_{0i}=\int d^2x [{\tilde \pi}_{\a}\p_iz_{\a}
+{\tilde \pi}^*_{\a}\p_iz^*_{\a}-2\t A^a_iB^a+2\T {\cal A}_ij_0]\eqno(4.9)$$
generate appropriate translation,
$$\{Z(x),{\tilde P}_i\}=\p_iZ(x)$$
$$\{A^a_k(x),{\tilde P}_i\}=\p_iA^a_k(x)\eqno(4.10)$$
just like $P_i^N$ (4.6).

So finally the corresponding expression of angular momentum can be written
as,
$$J^s=\int d^2x \e^{ij}x_i{\tilde T}_{0j}
=\int d^2x \e^{ij}x_i[{\tilde \pi}_{\a}\p_jz_{\a}
+{\tilde \pi}^*_{\a}\p_jz^*_{\a}-2\t A^a_jB^a+2\T {\cal A}_jj_0]\eqno(4.11)$$
$$J^N=\int d^2x [\e^{ij}x_i({\tilde \pi}_{\a}\p_jz_{\a}
+{\tilde \pi}^*_{\a}\p_jz^*_{\a}-\t \e^{kl}A^a_k\p_jA^a_l)-\t A^a_jA^a_j]
\eqno(4.12)$$
Just like the case the case of linear momentum, here too one can show that
both $J^s$ and $J^N$ generate appropriate spatial rotation,
$$\{Z(x),J^s\}=\{Z(x),J^N\}=\e^{ij}x_i\p_jZ(x)$$
$$\{A_k^a(x),J^s\}=\{A^a_k(x),J^N\}=\e^{ij}x_i\p_jA^a_k(x)+\e_{ki}A^{ai}(x)\eqno(4.13)$$
(Again the adjective ``appropriate" has been used in the sense, mentioned
above.)
However, they are not identical and the difference $J_f\equiv (J^s-J^N)$ is
given as,
$$J_f=\t \int d^2x \p_i[x_jA^{aj}A^{ai}-x^iA^a_jA^{aj}]
+2\T \int d^2x \e^{ij}x_i{\cal A}_jj_0 \eqno(4.14)$$
The first $\t$-dependent boundary term has occured earlier in [8,9], where
some of its properties were studied in detail. For example, it was noted that
this term is gauge invariant under only those gauge transformations which
tends to identity asymptotically[9]. As mentioned earlier, only the set
of such transformations constitute the gauge group of the model (2.1).
Thus $J_f$ is actually gauge invariant under the complete gauge group of
(2.1). To evaluate it in a rotationally symmetric
configuration therefore, one can make use of the radiation gauge
$(\p_iA_i^a=0)$ condition. (Clearly, this corresponds to one particular
sector of the theory). To this end, let us rewrite the Gauss constraint
(3.27) as,
$$j^a_0\approx 2\t B^a \eqno(4.15a)$$
with
$$j^a_0=ig({\tilde \pi}_{\a}(T^aZ)_{\a}-{\tilde
\pi}^*_{\a}(Z^{\dg}T^a)_{\a})\eqno(4.15b)$$
Note that the global $SU(2)$ invariance of the model (3.21) yields the
following conserved $3(=2+1)$-current
$$J^{a\mu}=ig[(D^{\mu}Z)^{\dg}T^aZ-Z^{\dg}T^a(D^{\mu}Z)] -
g\t\e^{\mu\nu\l}\e^{abc}A_{\nu}^bA_{\l}^c$$
$$+{\T g\over \pi}\e^{ij}{\cal A}_i[(Z^{\dg}T^a(D_jZ)+(D_jZ)^{\dg}T^aZ)\d^{\mu}_0-
(Z^{\dg}T^a(D_0Z)+(D_0Z)^{\dg}T^aZ)\d^{\mu}_j]\eqno(4.16)$$
satisfying $\p_{\mu}J^{\mu a} = 0$. The time component $J^{a0}$, when
simplified using (3.22) and (4.15), yields
$$J^{a0} = 2\t b^a\eqno(4.17a)$$
where
$$b^a = \p_1 A^a_2 - \p_2 A^a_1\eqno(4.17b)$$
is the `abelianized' part of the nonabelian magnetic field $B^a$ (2.5c).
This shows that the associated $SU(2)$ conserved charges
$Q^a(\equiv \int d^2x J^a_0)$ are related to the triplet of `abelianized'
fluxes $\Phi^a_{abelian} = \int d^2x b^a$ by
$$Q^a\approx 2\t \Phi^a_{abelian} \eqno(4.18)$$
rather then the flux of the nonabelian magnetic field $(\Phi=\int d^2x
B^a)$.

Proceeding as
in [8], we can write the following configuration of the $SU(2)$ gauge field,
$$A^a_i=-{Q^a\over 4\pi \t}\e_{ij}{x^j\over r^2}\eqno(4.19)$$
in the radiation gauge ($\p_iA^a_i=0$). Using this, one can easily show that
the first $\t$-dependent term in (4.14) yields ${Q^aQ^a\over 4\pi \t}$ and
the second $\T$-dependent term yields, following [13], $\T N^2$. So finally,
we have from (4.14),
$$J_f={Q^aQ^a\over 4\pi \t}+\T N^2 \eqno(4.20)$$
(Incidentally, it turns out that in gauge (4.19), $Q^a = \int d^2x J^a_0
= \int d^2x j^a_0$.)
We thus see that the classical expression of fractional angular momentum
contains two terms. One depends on the soliton number $N$ and the other on
the nonabelian charge $Q^a$. The former is just
as in the model, where Hopf term is coupled to NLSM [3,13]. On the other hand,
the latter is a typical nonabelian expression, as in [8]. It needs to be
mentioned here that for preservation of invariance of the action
(2.1) or (3.21) involving the nonabelian ($SU(2)$) CS term under a
homotopically nontrivial gauge transformation, the CS coefficient $\t$
is quantized  ($\t = n/8\pi$) with $n \in Z$ being an integer~[23].
In the next
section we shall compute $J_f$ (for $\T = 0$) for a physically different
sector of the theory.

\vskip 0.2in

5. {\bf Reduced phase space analysis of the model}

\vskip 0.1in

In this section we shall consider a reduced phase space version of the model
(2.1) associated with a different physical sector of the theory defined
by (3.12). At first we perform its Hamiltonian analysis in terms of the
surviving dynamical degrees of freedom. Upon using (3.12),
i.e.,
$$Z = \Pm{0 \cr 1}\eqno(5.1)$$
the degrees of freedom
corresponding to the $CP^1$ matter fields $Z$ get completely eliminated.
This configuration can be obtained by making use of the local $SU(2)$
gauge invariance of the model (2.1).
With this choice, we have broken the $SU(2)$ {\g} symmetry of the model
(2.1). Thus with the condition (5.1), the configuration space and hence
the phase space get reduced partially.
As will become apparent subsequently, this still leaves
a residual $U(1)$ symmetry. We shall obtain the generator of {\am} and
also the expression for {\fs} in this model. Before concluding this
section, we shall investigate whether one can have any topological
solitons, and hence the Hopf term in this case also.

The substitution of (5.1) in the model (2.1) yields the {\g} fixed
Lagrangian
$${\cal L} = {g^2 \over 4}A^{\a}_{\mu}A^{\mu\a} + \t\e^{\mu\nu\l}
(A^a_{\mu}\p_{\nu}A^a_{\l} + {g \over 3}
\e^{abc}A^a_{\mu}a^b_{\nu}A^c_{\l})\eqno(5.2)$$
where the group index $\a =1,2$. (Henceforth,
the first two Greek indices $\a$ and $\b$ will take
values $1$ or $2$ only. All the other symbols will retain their usual
meanings.) This form of the Lagrangian clearly shows that the original
$SU(2)$ symmetry could not be broken entirely. The $U(1)$ subgroup of
$SU(2)$ (see 2.34) survives as a {\g} symmetry. Thus $G^1_2$ and $G^2_2$
of (2.7) correspond to the broken generators of $SU(2)$, and $G^3_2$
corresponds to the surviving $U(1)$ symmetry. (Note, that we do not
bother about the other $U(1)$ symmetry generator $G_1$ (2.6) since it is
not an independent constraint (2.32).) Correspondingly, there are
mass-like terms for $A^1_{\mu}$ and $A^2_{\mu}$ in the Lagrangian (5.2),
whereas, $A^3_{\mu}$ remains massless. To understand better the survival
of the residual $U(1)$ symmetry even after making the gauge choice
(5.1), note that had we chosen a configuration of the $Z$-field as $Z(x)
= \Pm{0 \cr e^{i\phi(x)}}$, the Lagrangian would still have taken the
form (5.2). Actually, all these $Z$-field configurations correspond to
the same $M^a$ field configuration $(M^a = -\d^{a3})$, so that
$\stackrel{\ra}{M}$ points towards the south pole in the
isospin space of unit radius. The surviving $U(1)$ symmetry (2.34) is
the $SO(2)$ rotational symmetry around the $z$-axis. The situation is
somewhat analogous to the Higgs mechanism of the standard model where
broken symmetry generators provide masses for the gauge fields. The
constraint (2.2) on the $Z$-fields here plays the role of the vacuum
expectation value of the Higgs field in the standard model.

Coming to the Hamiltonian analysis of the model, the canonically
conjugate momenta corresponding to the surviving physical variables in
the reduced configuration space $(A^a_i, A^a_0)$ are given by
$$\pi^{ia} = {\d L \over \d \dot{A}^a_i} = \t \e^{ij}A^a_j ; \pi^{0a} =
{\d L \over \d \dot{A}^a_0} = 0\eqno(5.3)$$
The Legendre transformed Hamiltonian can be written as
$${\cal H} = - {1 \over 4} g^2 A^{\a}_{\mu}A^{\mu\a} - 2\t
A^a_0B^a\eqno(5.4)$$
It is apparent that $A^3_0$ is just a Lagrange multiplier enforcing the
Gauss constraint
$$G \equiv -2\t B^3 \approx 0\eqno(5.5)$$
The fact that $G$ generates the appropriate $U(1)$ {\g} {\tr} will be
demonstrated later. Now note that preservation of the pair of
constraints $\pi^{0\a} \approx 0$ (5.3) in time yield the following pair of
secondary constraints:
$$ {1\over 2} g^2 A^{\a}_0 + 2\t B^{\a} \approx 0\eqno(5.6)$$
It can be checked that the constraint $\pi^{0\a} \approx 0$ (5.3)
together with (5.6) form two pairs of
second class constriants which are strongly implemented by the following
DB's:
$$\{A^a_i(x), A^b_j(y)\} = {1 \over 2\t} \e_{ij}\d^{ab}
\d(x-y)\eqno(5.7a)$$
$$\{A^a_0(x), A^b_0(y)\} = 0\eqno(5.7b)$$
$$\{A^a_0(x), A^b_i(y)\} = {2\over g^2} \d^{a\a} \d^{b\a} \p^{(x)}_i
\d(x-y) - {2\over g} \d^{a\a} \e^{\a bc} A^c_i(x)\d(x-y)\eqno(5.7c)$$
This leaves the first class constraint (5.5)
$$G = -2\t \e^{ij}(\p_i A^3_j + {g\over 2} \e^{\a\b}A^{\a}_iA^{\b}_j)
\approx 0\eqno(5.8)$$
as the generator of the surviving $U(1)$ gauge symmetry with respect to
the DB's (5.7), yielding
$$\d A^3_i(x) = \int d^2y f(y)\{A^3_i(x), G(y)\} = \p_i f(x)\eqno(5.9)$$
as expected. Furthermore, the action of $G$ on the `massive' fields
$A^{\a}_{\mu}$ yields
$$\d A^{\a}_{\mu}(x) = \int d^2y f(y) \{A^{\a}_{\mu}(x), G(y)\} =
gf(x)\e^{\a\b}A^{\b}_{\mu}(x)\eqno(5.10)$$
which is just a rotation in the two-dimensional internal space spanned
by $\a$ and $\b$.

We now define the energy-momentum tensor $T^N_{\mu\nu}$ using the
Noether prescription, from which it follows that the expression of
momentum $P^N_j$ is given by
$$P^N_j = \int d^2x T^N_{0j} = -\t \int d^2x [\e^{ik}A^a_i(x)\p_j
A^a_k(x)]\eqno(5.11)$$
From here the Noether angular momentum $J^N$ is computed to be
$$J^N = -\t \int d^2x [\e^{ij}x_i\e^{kl}A^a_k\p_jA^a_l +
A^a_jA^a_j]\eqno(5.12)$$
It is easy to check that both the momentum $P^N_j$ (5.11) and \am $J^N$
(5.12) respectively generate appropriate translation defined in (4.6),
and appropriate rotation defined in (4.13) on the dynamical variables
$A^a_i$ and $A^{\a}_0$.

Next we write down the symmetric energy momentum tensor
$$T^s_{\mu\nu} = {1\over 2}g^2A^{\a}_{\mu}A^{\a}_{\nu} - {1\over
4}g^2g_{\mu\nu}A^{\a}_{\rho}A^{\a\rho}\eqno(5.13)$$
as obtained by functionally differentiating ther action (5.2) with
respect to the metric. (Note that the expression (5.13) can also be
obtained from (3.1) by substituting (5.1) in it.) Just as in (4.4),
here too the
expression of momentum obtained from (5.13) fails to generate
appropriate translation. It thus requires to be modified by adding a suitable
combination of first class constraint(s), here only (5.8):
$${\tilde T}_{\mu\nu} = T^s_{\mu\nu} + w_{\mu\nu}G\eqno(5.14)$$
It follows that by choosing $w_{0j} = A^3_j$, one gets the desired
expression for symmetric momentum
$${\tilde P}^s_j = \int d^2x {\tilde T}^s_{0j} = \int d^2x({g^2 \over 2}
A^{\a}_0 A^{\a}_j + GA^3_j)\eqno(5.15)$$
which generates the appropriate translations defined in (4.10) on
$A^a_i$ and $A^{\a}_0$. Upon using the relation (5.6), it follows that
(5.15) reduces to a simple expression
$${\tilde P}^s_j = -2\t \int d^2x B^aA^a_j\eqno(5.16)$$
From here the symmetric \am $J^s$ is given by
$$J^s = -2\t \int d^2x \e^{ij}x_iB^aA^a_j = -2\t \int d^2x \e^{ij}
\e^{lk} x_i A^a_j \p_l A^a_k\eqno(5.17)$$
which, again, generates appropriate rotations.

Before proceeding to compute the {\fs} given in this case by the
difference between $J^s$ (5.17) and $J^N$ (5.12), let us now consider
the expression for energy $E$ obtained by integrating $T^s_{00}$ (from
5.13) to get
$$E = \int d^2x T^s_{00} = \int d^2x {g^2 \over 4} [(A^{\a}_0)^2 +
(A^{\a}_i)^2]\eqno(5.18)$$
Clearly, if one demands that the energy $E$ should be finite, it is
necessary that $A^{\a}_{\mu}$ vanish at infinity. This is consistent
with the presence of masslike terms for the fields $A^{\a}_{\mu}$ in
Lagrangian (5.2).

The expression for {\fs} is given by
$$J_f = J^s -J^N = \t \int d^2x \p_i(x_jA^{j\a}A^{i\a} -
x^iA^{\a}_jA^{j\a} + x_jA^{j3}A^{i3} - x^iA^3_jA^{j3})\eqno(5.19)$$
which is identical to the first term of (4.14). The first two terms on
the right hand side correspond to boundary values of `massive' fields
$A^{\a}_i$, and hence vanish identically. One is thus left with
$$J_f = \t \int d^2x \p_i(x_jA^{j3}A^{i3} - x^i
A^3_jA^{j3})\eqno(5.20)$$
where $A^3_i$ is the surviving $U(1)$ gauge field.

As in the previous section, the global $U(1)$ invariance of the model
(5.2) yields the current
$$J^{\mu} = \t
g\e^{\mu\nu\l}\e^{\a\b}A^{\a}_{\nu}A^{\b}_{\l}\eqno(5.21)$$
satisfying $\p_{\mu}J^{\mu} = 0$. The corresponding conserved charge is
$$Q = \int d^2x J^0 = \t g\int d^2x
\e^{ij}\e^{\a\b}A^{\a}_iA^{\b}_j\eqno(5.22)$$
Using the Gauss constraint (5.8), $Q$ can be written, on the constraint
surface, entirely in terms
of the gauge field $A^3_i$ as
$$Q \approx 2\t\e^{ij}\int d^2x \p_iA^3_j\eqno(5.23)$$
It is easy to verify that the form of $Q$ (5.22), rather than (5.23),
generates the {\tr} (5.10)
$$\{A^{\a}_{\mu}(x), Q\} = g\e^{\a\b}A^{\b}_{\mu}\eqno(5.24)$$
This is expected since (5.23) is only a `weak' equality.

Now one can invoke the radiation {\g} condition ($\p_iA^3_i = 0$)
for this surviving $U(1)$ gauge field $A^3_i$, which yields the
following asymptotic form for it:
$$A^3_i = -{Q\over 4\pi\t}\e_{ij}{x^j \over r^2}\eqno(5.25)$$
Using this form for the {\g} field, the fractional {\am} $J_f$ (5.20) is
computed to be
$$J_f = {Q^2 \over 4\pi\t}\eqno(5.26)$$
One can see clearly that this is different from (4.20) in the absence of
the Hopf term ($\T = 0$). This difference stems from the fact that the
radiation gauge condition $(\p_iA^a_i = 0)$ and the condition $(Z =
\pmatrix{0 \cr 1})$ (5.1) correspond to different physical sectors of
the theory.
conditions used. As we have seen, these conditions are
associated with different asymptotic properties of the CS fields here, and this
in turn has different consequences on the physics of the system in this
case.

Finally, let us consider the effect of addition of the Hopf term, if
any, to the model (5.2). For this we have to at first look for a
solitonic configuration in this model. To that end, note that the
expression for energy $E$ (5.18) can be written as
$$E = {g^2 \over 4} \int d^2x [(A^1_0)^2 + (A^2_0)^2 + (A^1_2 \pm
A^2_1)^2 + (A^1_1 \mp A^2_2)^2 \pm
2\e^{ij}\e^{\a\b}A^{\a}_iA^{\b}_j]\eqno(5.27)$$
(Again, the expression (5.27) can also be obtained from (3.3a) by making
the substitution $Z = \Pm{0 \cr 1}$ (5.1) in it).
Using the expression of Noether charge (5.22), this yields the following
Bogomol'nyi type inequality:
$$E \ge {g|Q| \over 2\t}\eqno(5.28)$$
The saturation condition corresponds to the static field configurations
satisfying
$$A^1_0 = A^2_0 = 0$$
$$A^1_2 \pm A^2_1 = 0\eqno(5.29)$$
$$A^1_1 \mp A^2_2 = 0$$
This describes the solitonic configuration in the model. But note that
the local minimum of energy is now given by the Noether charge $Q$ which
here plays the role of topological charge. With condition that
$A^{\a}_{\mu}$ vanish at infinity, the two dimensional plane gets
effectively compactified to the $2$-sphere $S^2$, and the `weak'
equality (5.23) allows one to identify $Q$, albeit `weakly', with the
first Chern class. Thus the field configuration (5.29) can be identified
`weakly' with a topological soliton [4]. However, it turns out that this
solitonic configuration is a rather trivial one. This can be seen
clearly from (5.28). Note that
the current $J^{\mu}$ (5.21) has vanishing spatial
components $(J^i = 0)$, which in turn implies that the Hopf term
(3.20) vanishes~[13].

\vskip 0.2in

6. {\bf Conclusions}

\vskip 0.1in

In this paper, we have carried out a detailed classical Hamiltonian analysis of
the gauged $CP^1$ model of Cho and Kimm[19]. This model is obtained by
gauging the global $SU(2)$ group of the $CP^1$ model which is already a
$U(1)$ gauge theory. We find that contrary to our expectation, the gauge
group of this gauged $CP^1$ model turns out to be only $SU(2)$, rather
than $SU(2) \times U(1)$.
As was shown in [19], the
model has got its own solitons, the very existence of which depends
crucially on the presence of $SU(2)$ CS term. These solitons are
somewhat more general then that
of NLSM[10]. We use the adjective ``general'' to indicate that these
solitons can be characterised by ${\pi_2(CP^1)}=Z$ only for the gauge
$A_i^a=0$ (Note that $A_i^a$ is a pure gauge). One can make a topology
changing  transformation  and thereby make $A^a_i\neq 0$, without
changing the soliton number. However, such a transformation does not
become constant asymptotically, and therefore does not belong to the
group of invariance, viz., gauge group of the model.
We then constructed the Hopf term associated
to these solitons and again carried out the Hamiltonian analysis of the
model(3.21), obtained by adding Hopf term to (2.1), to find that
the symplectic structure and the structure of the
constraints undergo essentially no modification, despite the fact that
the form of the momenta variables conjugate to $z_{\a}$ and $z^*_{\a}$
undergo changes. We then calculated the fractional angular momentum
by computing the difference between $J^s$ and $J^N$, the expressions
of angular momenta obtained from the symmetric expression of energy-momentum
tensor and the one obtained through Noether's prescription respectively.
We find that this fractional angular momentum consists of two pieces,
one is given in terms of the soliton number and the other is given in terms
of the nonabelian ($SU(2)$) charge. In absence of the
Hopf term $(\T =0)$ (i.e. for the model (2.1)), only this latter term will
contribute. Again as in [8], this term can be shown to consist of two
pieces, one which involves a direct product in the isospin space and
characterises a typical nonabelian feature, while the other contains
the abelian charge defined in a nonabelian theory.

Subsequently, by making use of the local $SU(2)$ gauge invariance of the
model, we can essentially eliminate all the degrees of freedom
associated with the $Z$ fields. In fact, with the choice $Z = \Pm{0 \cr
1}$ (5.1), the $CP^1$ fields are frozen out, and one is considering a
distinct physical sector from the one considered earlier.
With this  choice, we perform a (partially)
reduced phase space Hamiltonian analysis of the resultant model. Here we
see that the $SU(2)$ symmetry is only partly broken with a residual abelian
$U(1)$ symmetry. Correspondingly, masslike terms are generated for the
two $A^{\a}_{\mu}$ fields, whereas, $A^3_i$ survives as a $U(1)$ gauge
field. This situation is somewhat akin to the standard model where
the {\g} symmetry is partially broken and corresponding mass
terms for the gauge fields are generated by the Higgs mechanism. The
role of the vacuum expectation value of the Higgs field is played here
by the constraint on the magnitude of the $Z$-fields. We next calculate the
expressions of fractional spin through the two angular momenta
in our model. We find that the value of
{\fs} depends in this case purely on the abelian charge of the surviving
$U(1)$ symmetry. This result of {\fs} is different from the one obtained
earlier using the radiation gauge condition. This indicates that different
physical sectors of the theory are associated with different fractional
angular momentum.
Finally, we use a Bogomol'nyi type inequality to find
the static minimum energy configuration for this model. This defines a
solitonic configuration for the model. The resultant
solitonic charge turns out to be equal to the Noether charge here. The
solitonic configuration however, turns out to be of a trivial nature, in
the sense that the corresponding Hopf term vanishes.

\pagebreak

{\bf References}

\begin{enumerate}

\item F.Wilczek (Ed.) ``Fractional Statistics and Anyonic Superconductivity'',
(World Scientific, Singapore, 1990)

\item S.Forte, Rev.Mod.Phys.{\bf 64}(1992)193.

\item F.Wilczek and A.Zee, Phys.Rev.Lett.{\bf 51}(1983)2250.

\item A.P.Balachandran, G.Marmo, B.S.Skagerstam and A.Stern, ``Classical
Topology and Quantum States'', (World Scientific, Singapore,1991).

\item C.R.Hagen, Ann.Phys.(N.Y.){\bf 157}(1984)342.

\item P.Panigrahi, S.Roy and W.Scherer, Phys.Rev.Lett. {\bf 61}(1988)2827.

\item R.Banerjee, Phys. Rev. Lett. {\bf 69} (1992) 17; Phys. Rev. {\bf
D41} (1993) 2905; Nucl. Phys. {\bf B390} (1993) 681; R.Banerjee and
B.Chakraborty, Phys. Rev. {\bf D49} (1994) 5431.

\item R.Banerjee and B.Chakraborty, Ann.Phys.(N.Y.){\bf 247}(1996)188.

\item B.Chakraborty, Ann. Phys. (N.Y.) {\bf 244} (1995) 312; B.
Chakraborty and A. S. Majumdar, Ann. Phys. {\bf 250} (1996) 112.

\item R.Rajaraman, ``Solitons and Instantons'' (North Holland, 1982);
C.Rebbi and G.Soliani, ``Solitons and Particles'' (World Scientific,
Singapore, 1984); V.A.Novikov, M.A.Shifman, A.I.Vainshtein and
V.I.Zhakharov, Phys. Rep. {\bf 116} (1984) 103.

\item Y.S.Wu and A.Zee, Phys.Lett. {\bf 147B}(1984)325.

\item B.Chakraborty and T.R.Govindarajan, Mod.Phys.Lett.{\bf A12}(1997)619.

\item M.Bowick,D.Karabali and L.C.R.Wijewardhana, Nucl.Phys.{\bf B271}(1986)417.

\item J.Zinn-Justin,``Quantum Field Theory and Critical Phenomena''
(Clarendon Press,Oxford, 1990); E.Fradkin,``Field Theory of Condensed matter
systems''(Addison-Wesley,1991).

\item L.Landau and E.Lifshitz, Phys. A. {\bf 8} (1935) 153; A. Kosevich,
B. Ivanov and A. Kovalev, Phys. Rep. {\bf 194} (1990) 117.

\item R.Banerjee and B.Chakraborty, Nucl.Phys.{\bf B449}(1995)317.

\item H.Kobayashi, I.Tsutsui and S.Tanimura,``Quantum Mechanically Induced
Hopf Term in the $O(3)$ non-linear sigma model'', hep-th/9705183.

\item G.Nardelli, Phys.Rev.Lett. {\bf 73}(1994)2524.

\item Y.Cho and K.Kimm, Phys.Rev.{\bf D52}(1995)7325.

\item P.A.M.Dirac, ``Lectures on Quantum Mechanics'',  Belfar Graduate School of
Science, Yeshiva University, New York, 1964; A.Hanson, T.Regge and
C.Teitelboiml ``Constraint Hamiltonian Analysis'', (Academia Nazionale
dei Lincei, Rome, 1976).

\item L.Faddeev and R.Jackiw, Phys.Rev.Lett.{\bf 60}(1988)1692.

\item R.Banerjee, Phys.Rev.{\bf D49}(1994)2133.

\item S.Deser, R.Jackiw and S.Templeton, Ann. Phys. (N.Y.) {\bf 140}
(1982) 372.

\end{enumerate}

\end{document}